\documentclass[12pt]{article}
\usepackage{cite}
\usepackage{CJK}
\usepackage{amssymb}
\usepackage{amsmath}
\usepackage{enumerate}
\usepackage{graphicx}
\usepackage{amsfonts}
\usepackage{url}
\usepackage{bm}

\usepackage{pifont}
\usepackage{epsfig,subfigure,dsfont,amsthm,amsbsy,mathrsfs,amscd}

\def\be{\begin{equation}}
\def\ee{\end{equation}}
\def\ba{\begin{array}}
\def\ea{\end{array}}
\input amssym.def
\newcommand\btd{\raise 2pt \hbox{$\hat\bigtriangledown$}\hskip 1.5pt}
\newcommand\bt{\raise 2pt \hbox{$\bigtriangledown$}\hskip 1.5pt}
\begin{document}
 \title{\large\bf Improved Separability Criteria Based on Bloch Representation of Density Matrices}
\author{Shu-Qian Shen$^\dag$, Juan Yu$^\dag$, Ming Li$^\dag$$^\ast$ and Shao-Ming
Fei$^\sharp$$^\flat$\\[10pt]
\footnotesize \small $^\dag$College of the Science, China University
of Petroleum,\\
\small Qingdao 266580, P. R. China\\
\footnotesize
\small $^\sharp$School of Mathematical Sciences, Capital Normal University,\\
\small Beijing 100048, P. R. China\\
\small $^\flat$Max-Planck-Institute for Mathematics in the Sciences,\\
\small Leipzig 04103, Germany}
\date{}

\maketitle

\centerline{$^\ast$ Correspondence to liming3737@163.com}
\bigskip

\begin{abstract}

The correlation matrices or tensors in the Bloch representation of density matrices are encoded with entanglement properties. In this paper, based on the Bloch representation of density matrices, we give some new separability criteria for bipartite and multipartite quantum states. Theoretical analysis and some examples show that the proposed criteria can be more efficient than the previous related criteria.
\end{abstract}
\bigskip

Quantum entanglement is a fascinating phenomenon in quantum physics. It can be seen as a physical resource like energy with applications from quantum teleportation to quantum cryptography \cite{applications,applications1,applications2,applications3,applications4}. In the last years, much work has been devoted to understanding entanglement, but there are still many problems unsolved. One of them is to determine whether a given quantum state is entangled or separable. This problem is extremely difficult to solve, and has been proved as a nondeterministic polynomial-time hard problem \cite{Gurvits2003-1}.  Nevertheless, a variety of operational criteria for separability of quantum states have been proposed in the last decades. Among them are the positive partial transpose (PPT) criterion or Peres-Horodecki criterion \cite{ppt,ppt1}, realignment criteria \cite{realignment1,realignment11,realignment2,realignment21,realignment22}, covariance matrix criteria \cite{covariance,covariance1,covariance-multi} and so on; see, e.g., \cite{survey,survey1} for a comprehensive survey.

The Bloch representation \cite{Bloch,Bloch1,Bloch2} of density matrices stands as an important role in quantum information. The correlation matrices or tensors in the Bloch representation are encoded with entanglement properties \cite{correlation-geometry,correlation-geometry1}, which can be exploited to study quantum entanglement. In \cite{Vicente2007}, by making use of correlation matrices, Vicente obtained the correlation matrix criterion for bipartite quantum states, which can be more efficient than the PPT criterion \cite{ppt,ppt1} and the computable cross norm or realignment (CCNR) criterion \cite{realignment1,realignment11} in many different situations. After that, this criterion was used to give the analytical lower bounds for the entanglement measures: concurrence and tangle \cite{Vicente-concurrence-tangle,Vicente-concurrence-tangle1}, which are good supplement to the lower bounds based on PPT and CCNR criteria. By the matricizations of tensors, the correlation matrix criterion was generalized to detect non-full-separability of multipartite states \cite{Hassan}. Later, this multipartite criterion was extended and improved to be a much more general case \cite{multi-general}. Meanwhile, by the standard tensor norm and the norms of matricizations of tensors, some genuine entanglement conditions were derived. In  \cite{correlation-geometry,correlation-geometry1}, some simple geometrical methods based on correlation tensors were presented to detect various multipartite entanglement. By bounding tensor norms for partially separable states and states of limited dimension, Kl\"{o}ckl and Huber \cite{k-separable} studied the detection of multipartite entanglement in an experimentally feasible way. In many cases, only few definite measurements are needed. Recently, Li et al. \cite{Liming} presented some separability criteria under the combination of correlation matrices and the Bloch vectors of reduced density matrices, which can be stronger than the correlation matrix criterion \cite{Vicente2007} by examples.

 This paper is further devoted to an investigation of entanglement detection in terms of Bloch representations of density matrices. On the one hand, by adding some parameters, a more general separability criterion for bipartite states is presented, which can outperform the corresponding criteria given in \cite{Vicente2007,Liming}. On the another hand, the presented bipartite separability criterion is extended to the multipartite case. An example shows that the new multipartite separability criterion can be better than the corresponding criteria obtained in \cite{Hassan,multi-general,Liming}.

\medskip
\noindent{\bf Results}
\medskip

{\sf Separability criteria for bipartite states.}~Let $\lambda_i^{(d)},i=1,2,\cdots,d^2-1$ be the traceless Hermitian generators of $SU(d)$ satisfying the orthogonality relation $\text{Tr}(\lambda_i^{(d)}\lambda_j^{(d)})=2\delta_{ij}$. Then any state $\rho$ in $\mathbb{C}^{d_1}\otimes \mathbb{C}^{d_2}$ can be represented as \cite{Bloch2}
\begin{equation}
\label{Bloch}
\rho=\frac{1}{d_1d_2} \left(I_{d_1}\otimes I_{d_2}+\sum\limits_{i=1}^{d_1^2-1}r_i\lambda_i^{(d_1)}\otimes I_{d_2}+\sum\limits_{j=1}^{d_2^2-1}s_j I_{d_1}\otimes \lambda_j^{(d_2)}+
\sum\limits_{i=1}^{d_1^2-1}\sum\limits_{j=1}^{d_2^2-1}t_{ij}\lambda_i^{(d_1)}\otimes \lambda_j^{(d_2)}
\right),
\end{equation}
where $I_{d}$ denotes the $d\times d$ identity matrix,
\begin{equation}
\label{eq:cm}
r_i=\frac{d_1}{2}\text{Tr}(\rho\lambda_{i}^{(d_1)}\otimes I_{d_2}),~
s_j=\frac{d_2}{2}\text{Tr}(\rho I_{d_1}\otimes \lambda_{j}^{(d_2)}),~
t_{ij}=\frac{d_1d_2}{4}\text{Tr}(\lambda_i^{(d_1)}\otimes \lambda_j^{(d_2)}).
\end{equation}
Denote by $||\cdot||_{\text{tr}}$, $||\cdot||_{2}$ and $E_{p\times q}$ the trace norm (the sum of singular values), the spectral norm (the maximum singular value) and the $p\times q$ matrix with all entries being 1, respectively. By defining $r=(r_1,\cdots,r_{d_1^2-1})^t$, $s=(s_1,\cdots,s_{d_2^2-1})^t$ and $T=(t_{ij})$, we construct the following matrix
\begin{equation}
\mathcal{S}_{\alpha,\beta}^{m}(\rho)=\left( {{\begin{array}{*{20}c}
\alpha\beta E_{m\times m}& {\beta \omega_{m}(s)^t}  \\[1mm]
 {\alpha \omega_{m} (r)}  &T \\
\end{array} }} \right),
\end{equation}
where $\alpha$ and $\beta$ are nonnegative real numbers, $m$ is a given natural number, $t$ stands for transpose, and for any column vector $x$,
\begin{equation}
 \omega_{m} (x) = \underbrace {\left( {\begin{array}{*{20}{c}}
   {x} &  \cdots  & {x}  \\
\end{array}} \right)}_{m\;\; columns}.
\end{equation}
Using $\mathcal{S}_{\alpha,\beta}^{m}(\rho)$, we can get the following separability criterion for bipartite states.

{\textbf{Theorem 1.}} If the state $\rho $ in $\mathbb{C}^{d_1}\otimes \mathbb{C}^{d_2}$ is separable, then
\begin{equation}
\label{eq:thm211}
||\mathcal{S}_{\alpha,\beta}^{m}(\rho)||_{\text{tr}}\le \frac{1}{2}\sqrt{(2m\beta^2+d_1^2-d_1)(2m\alpha^2+d_2^2-d_2)}.
\end{equation}

See Methods for the proof of Theorem 1.

When $\alpha$ and $\beta$ are chosen to be $0$, Theorem 1 reduces to the correlation matrix criterion in \cite{Vicente2007}: if $\rho $ in $\mathbb{C}^{d_1}\otimes \mathbb{C}^{d_2}$ is separable, then
\begin{equation}
\label{eqVi}
||T||_{\text{tr}}\le \frac{1}{2}\sqrt{(d_1^2-d_1)(d_2^2-d_2)}.
\end{equation}
If we choose $\alpha=\beta=m=1$, then Theorem 1 becomes the separability criterion given in \cite[Corollary 2]{Liming}: any separable state $\rho $ in $\mathbb{C}^{d_1}\otimes \mathbb{C}^{d_2}$ must satisfy
\begin{equation}
\label{eqLi} {\left\| {\left( {\begin{array}{*{20}{c}}
   1 & {{s^t}}  \\
   r & T  \\
\end{array}} \right)} \right\|_{\text{tr}}} \le  \frac{1}{2}\sqrt{(2+d_1^2-d_1)(2+d_2^2-d_2)}.\end{equation}
For simplicity, we call these criteria in (\ref{eqVi}) and (\ref{eqLi}) the V-B and L-B criteria, respectively.
The following result can help us find that our separability criterion from Theorem 1 is stronger than the V-B and L-B criteria.

\textbf{Proposition 1.} If $\alpha$ and $\beta$ are selected to satisfy
\begin{equation}
\label{eq:condition}\alpha\sqrt{d_1(d_1-1)}=\beta\sqrt{d_2(d_2-1)},
\end{equation}
 then Theorem 1 becomes more effective when $m$ gets larger.

See Methods for the proof of Proposition 1.

From Proposition 1, Theorem 1 with the condition (\ref{eq:condition}) is stronger than the V-B criterion.

For the case $d_1=d_2$ and $\alpha=\beta$, it follows from Proposition 1 that Theorem 1 is more
efficient when $ m$ gets larger. In particular, Theorem 1 is better than the L-B criterion, and the L-B criterion is better than the V-B criterion. For the case $d_1\neq d_2$, let us consider the following example. The following $2\times 4$ bound entangled state is due to \cite{Horodecki1997}:
\begin{equation}\rho=\frac{1}{{7b + 1}}\left( {\begin{array}{*{20}{c}}
   b & 0 & 0 & 0 & 0 & b & 0 & 0  \\
   0 & b & 0 & 0 & 0 & 0 & b & 0  \\
   0 & 0 & b & 0 & 0 & 0 & 0 & b  \\
   0 & 0 & 0 & b & 0 & 0 & 0 & 0  \\
   0 & 0 & 0 & 0 & {\frac{1}{2}(1{{ + b)}}} & 0 & 0 & {\frac{1}{2}\sqrt {1 - {b^2}} }  \\
   b & 0 & 0 & 0 & 0 & b & 0 & 0  \\
   0 & b & 0 & 0 & 0 & 0 & b & 0  \\
   0 & 0 & b & 0 & {\frac{1}{2}\sqrt {1 - {b^2}} } & 0 & 0 & {\frac{1}{2}(1{{ + b)}}}  \\
\end{array}} \right),
\end{equation}
where $0<b<1$. To verify the efficiency of the present criteria, we consider the state
\begin{equation}
\rho_x=x|\xi\rangle\langle\xi|+(1-x)\rho,
\end{equation}
where $|\xi\rangle=\frac{1}{\sqrt{2}}(|00\rangle+|11\rangle)$.
For simplicity, we choose
\begin{equation}
\alpha=\sqrt{\frac{2}{d_1(d_1-1)}},~\beta=\sqrt{\frac{2}{d_2(d_2-1)}},~m=1.
\end{equation}
Then Theorem 1 can detect the entanglement in $\rho_x$ for $0.2235\le x\le 1$, while
the V-B criterion and L-B criterion can only detect the entanglement in $\rho_x$ for
$0.2293\le x\le 1$ and $0.2841\le x\le 1$, respectively.
Thus, Theorem 1 is better than the V-B and L-B criteria.

\medskip
{\sf Separability criteria for multipartite states.}~Let $\mathcal{S}$ be an $f_1\times \cdots \times f_N$ tensor, $A$ and $\bar A$ be two nonempty
subsets of $\{1,\cdots,N\}$ satisfying $A\bigcup \bar A=\{1,\cdots,N\}$. Then we denote
by $\mathcal{S}^{A|\bar A}$ the $A, \bar A$ matricization of $\mathcal{S}$; see \cite{multi-general}
for detail. This matricization is a generalization of mode-$n$ matricization in the multilinear algebra \cite{multilinear}.

For any state $\rho$ in $\mathbb{C}^{d_1}\otimes \cdots \otimes \mathbb{C}^{d_N}$, we import a natural number $m$ and nonnegative real parameters $\alpha_1,\cdots,\alpha_N$, and define
\begin{equation}
\delta _{{k_i}}^{({d_i})} = \left\{ \begin{array}{l}
 \frac{{2{\alpha _i}}}{{{d_i}}}{I_{{d_i}}},\quad 1 \le {k_i} \le m, \\[2mm]
 \lambda _{{k_i-m}}^{({d_i})},\quad m+1 \le {k_i} \le d_i^2 +m- 1,
 \end{array} \right. i=1,\cdots,N.
 \end{equation}
We define the tensor $\mathcal{W}_{\alpha_1,\cdots,\alpha_N}^{(m)}(\rho)$ with elements
\begin{equation}
w_{k_1\cdots k_N}=\frac{d_1\cdots d_N}{2^N}\text{Tr}(\rho \delta_{k_1}^{(d_1)}\otimes\cdots\otimes \delta_{k_N}^{(d_N)}),~ 1\le k_i\le d_i^2+m-1.
\end{equation}
Clearly, if $m=0$, the tensor $\mathcal{W}_{\alpha_1,\cdots,\alpha_N}^{(m)}(\rho)$ reduces to the correlation
tensor in \cite{Hassan}. When $m=\alpha_1=\cdots=\alpha_N=1$, the tensor
$\mathcal{W}_{\alpha_1,\cdots,\alpha_N}^{(m)}(\rho)$ becomes the tensor with a constant multiple in \cite{Liming}.

 An $n$ partite sate $\rho$ in
$ \mathbb{C}^{d_1}\otimes \cdots\otimes \mathbb{C}^{d_n}$ is (fully) separable  \cite{Werner1989} if and only if it can be written in the form
\begin{equation}
\rho=\sum\limits_i p_i\rho_i^1\otimes \cdots\otimes \rho_i^n,
\end{equation}
 where the probabilities $p_i> 0,~ \sum\nolimits_i p_i=1$, and $\rho_i^1,\cdots,\rho_i^n$ are pure states of the subsystems.

In the following, we give the full separability criterion based on  $\mathcal{W}_{\alpha_1,\cdots,\alpha_N}^{(m)}$.

\textbf{Theorem 2.} If the state $\rho$ in $\mathbb{C}^{d_1}\otimes \cdots \otimes \mathbb{C}^{d_N}$ is fully separable, then, for any subset $A$ of $\{1,\cdots,N\}$, we have
\begin{equation}
\left\|\left(\mathcal{W}_{\alpha_1,\cdots,\alpha_N}^{(m)}(\rho)\right)^{A|\bar A}\right\|_{\text{tr}}\le \mathop \Pi \limits_{k = 1}^N
\sqrt{\frac{1}{2}(2m\alpha_k^2+d_k^2-d_k)}.
\end{equation}

See Methods for the proof of Theorem 2.

For the case $\alpha_1=\cdots=\alpha_N=0$, Theorem 2 reduces to the criterion given in \cite[Theorem 4]{multi-general}, which has an important improvement on the corresponding criterion given in \cite{Hassan}. If $\alpha_1=\cdots=\alpha_N=1$ and $m=1$, then Theorem 2 becomes \cite[Corollary 3]{Liming}. For simplicity, we call these criteria in \cite{multi-general}, \cite{Hassan} and \cite{Liming} V-M, H-M and L-M criteria, respectively.
In the following we give a tripartite example to demonstrate the efficiency of Theorem 2. Consider a perturbation of the tripartite GHZ state \cite{covariance-multi}:
\begin{equation}
|\phi'_{GHZ}\rangle=\frac{1}{\gamma}(|000\rangle+\epsilon |110\rangle+|111\rangle),
\end{equation}
where $\epsilon$ is a given real parameter, and $\gamma$ denotes the normalization. We consider the mixture of this state with the maximally mixed state:
\begin{equation}
\rho_{GHZ}^{'x}=\frac{1-x}{8}I_8+ x|\phi'_{GHZ}\rangle \langle \phi'_{GHZ}|.
\end{equation}

In the tripartite case, the V-M criterion is equivalent to the H-M criterion obviously.
By taking $m=1$ and $\alpha_1=\alpha_2=\alpha_3=0.1$,
Table 1 displays  the detection results with different values of $\epsilon$.
Clearly, Theorem 2 is more efficient than the V-M, H-M and L-M criteria.

\begin{table}[htbp]
\centering \begin{tabular} {c|c|c|c}\hline
{$\epsilon$}&V-M (H-M) criteria & L-M criteria &Theorem 2
 \\ \hline
 {$0$}& $0.3536\le x\le 1$& $0.4118\le x\le 1$& $0.3307\le x\le 1$
 \\ \hline
 {$10^{-5}$}& $0.3536\le x\le 1$& $0.4118\le x\le 1$& $0.3307\le x\le 1$
 \\ \hline
 {$10^{-1}$}& $0.3424\le x\le 1$& $0.4118\le x\le 1$&  $0.3281\le x\le 1$
 \\ \hline
 {$1$}& $0.3274\le x\le 1$& $0.4256\le x\le 1$&  $0.3243\le x\le 1$
 \\ \hline
\end{tabular} \caption{\emph{Entanglement conditions of $\rho_{GHZ}^{'x}$ with different values of $\epsilon$ from
the V-M \emph{(}H-M\emph{) }criterion, the L-M criterion and Theorem 2 with $\alpha_1=\alpha_2=\alpha_3=0.1$ and $ m=1$.
}}
\label{tab1}
\end{table}

\medskip
\noindent{\bf Discussions}
\medskip

Correlation matrices or tensors in the Bloch representation of quantum states contain the information of
entanglement of the quantum states. Based on the Bloch representation of quantum states, we have given some new
separability criteria including the V-B, L-B, V-M, H-M and L-M criteria as special cases. For bipartite cases,
by choosing some special parameters involved, our criteria are stronger than the V-B and L-B criteria.
For multipartite cases, by a simple example it has been also shown that our criterion can be more efficient than the V-M, H-M and L-M criteria.

Nevertheless, the problem of how to choose the involved parameters such that Theorems 1-2 can detect more entangled states needs to be further studied in the future. In the Bloch representation  (\ref{Bloch}), the traceless Hermitian generators of $SU(d)$ come from Gell-Mann matrices. But this is by far not the only possible choice. Maybe the new basis of observables \cite{generators} constructed from Heisenberg-Weyl operators can be used to obtain better separable criteria, since the Heisenberg-Weyl based observables can outperform the canonical basis of generalized Gell-Mann operators in entanglement detection \cite{generators}. Thus, this problem is worth studying in the coming days.

It should be noted that the separability criteria Theorems 1-2 presented in \cite{Liming} for bipartite and multipartite states
are at most as good as the corresponding V-B, L-B, V-M and L-M criteria, respectively.
For example, set $\tilde T=(\tilde t_{kl})=\mathcal{S}_{1,1}^{1}(\rho)$. It was shown by \cite[Theorem 1]{Liming} that any separable state $\rho $ in $\mathbb{C}^{d_1}\otimes \mathbb{C}^{d_2}$ satisfies
\begin{align}
&\label{eqLi1} \left|\sum\limits_{kl}m_{kl}\tilde t_{kl}\right|\le \frac{1}{2}\sqrt{(2+d_1^2-d_1)(2+d_2^2-d_2)}||M||_2,
\end{align}
where  $M=(m_{ij})$ is any real $d_1^2\otimes d_2^2$ matrix. From (\ref{eqLi1}) and \cite{Watrous2011}, we get
\begin{align}
& \max\limits_{M\neq 0}\frac{\left|\sum\limits_{kl}m_{kl}\tilde t_{kl}\right|}{||M||_2}=
\max\limits_{M\neq 0}\frac{\left|\text{Tr}(M^t\tilde T)\right|}{||M||_2}=||\tilde T||_{\text{tr}}
\le \frac{1}{2}\sqrt{(2+d_1^2-d_1)(2+d_2^2-d_2)},
\end{align}
which implies that the L-B criterion is at least as good as the criterion (\ref{eqLi1}). Other cases can be proved similarly.

\medskip
\noindent{\bf Methods}
\medskip

{\sf Proof of Theorem 1.}~ Since $\rho$ is separable, from \cite[(17)]{Vicente2007}, it follows that there exist vectors $u_i\in \mathbb{R}^{d_1^2-1}$ and $v_j\in \mathbb{R}^{d_2^2-1}$ such that
\begin{equation}
T=\sum\limits_{i}p_iu_iv_i^t,~r=\sum\limits_{i}p_iu_i,~ s=\sum\limits_{i}p_iv_i,
\end{equation}
where
\begin{align}
&||u_i||_2=\sqrt{\frac{d_1(d_1-1)}{2}},~||v_j||_2=\sqrt{\frac{d_2(d_2-1)}{2}},~p_i\ge 0,~\sum\limits_{i}p_i=1.
\end{align}
Thus, the matrix $\mathcal{S}_{\alpha,\beta}^{m}(\rho)$ can be written as
\begin{align}
\mathcal{S}_{\alpha,\beta}^{m}(\rho)&=\sum\limits_i p_i \left( {{\begin{array}{*{20}c}
\alpha\beta E_{m\times m}& {\beta \omega_{m}(v_i)^t}  \\
 {\alpha \omega_{m} (u_i)}  &u_iv_i^t \\
\end{array} }} \right)\nonumber\\
&=\sum\limits_i p_i {\left( {\begin{array}{*{20}{c}}
   {\beta {E_{m \times 1}}}  \\
   {{u_i}}  \\
\end{array}} \right)\left( {\begin{array}{*{20}{c}}
   {\alpha {E_{1 \times m}}} & {v_i^t}  \\
\end{array}} \right)}
:=\sum\limits_i p_i \bar u_i\bar v_i^t,
\end{align}
and then
\begin{align}
||\mathcal{S}_{\alpha,\beta}^{m}(\rho)||_{\text{tr}}&\le \sum\limits_i p_i ||\bar u_i\bar v_i^t||_{\text{tr}}
=\sum\limits_i p_i ||\bar u_i||_2||\bar v_i||_2\nonumber\\
&=\frac{1}{2}\sqrt{(2m\beta^2+d_1^2-d_1)(2m\alpha^2+d_2^2-d_2)},
\end{align}
where we have used the following equality, for any vectors $|a\rangle$ and $|b\rangle$,
\begin{equation}
\label{eq:equality} |||a\rangle \langle b|||_{\text{tr}}=|||a\rangle||_2|||b\rangle||_2.
\end{equation}
\hfill \rule{1ex}{1ex}

{\sf Proof of Proposition 1.}~For any state $\rho$, from \cite[Lemma 1]{Vicente2007}, we get
\begin{equation}
\label{eq:pro221}||\mathcal{S}_{\alpha,\beta}^{m+1}(\rho)||_{\text{tr}}\ge \alpha\beta+||\mathcal{S}_{\alpha,\beta}^{m}(\rho)||_{\text{tr}}.
\end{equation}
If the inequality from (\ref{eq:thm211}),
\begin{equation}
||\mathcal{S}_{\alpha,\beta}^{m+1}(\rho)||_{\text{tr}}\le \frac{1}{2}\sqrt{(2(m+1)\beta^2+d_1^2-d_1)(2(m+1)\alpha^2+d_2^2-d_2)},\end{equation}
holds, then from (\ref{eq:pro221}) we have
\begin{align}
||\mathcal{S}_{\alpha,\beta}^{m}(\rho)||_{\text{tr}}&\le  \frac{1}{2}\sqrt{(2(m+1)\beta^2+d_1^2-d_1)(2(m+1)\alpha^2+d_2^2-d_2)}-\alpha\beta\nonumber\\[1mm]
& =\left\| {\left( {\begin{array}{*{20}{c}}
   {\beta {E_{(m+1) \times 1}}}  \\[1mm]
   {\sqrt{\frac{d_1(d_1-1)}{2}}}  \\[1mm]
\end{array}} \right)\left( {\begin{array}{*{20}{c}}
   {\alpha {E_{1 \times (m+1)}}} & {\sqrt{\frac{d_2(d_2-1)}{2}}}  \\[1mm]
\end{array}} \right)} \right\|_{\text{tr}}-\alpha\beta\nonumber\\[1mm]
\label{eq:pro222}&={\left\| {\begin{array}{*{20}{c}}
   {\alpha \beta {E_{(m+1) \times (m+1)}}} & {\beta \sqrt {\frac{{{d_2}({d_2} - 1)}}{2}} E_{(m+1)\times 1}}  \\[1mm]
   {\alpha \sqrt {\frac{{{d_1}({d_1} - 1)}}{2}} E_{1\times (m+1)}} & {\sqrt {\frac{{{d_1}{d_2}({d_1} - 1)({d_2} - 1)}}{4}} }  \\[1mm]
\end{array}} \right\|_{\text{tr}}}-\alpha\beta\nonumber\\[1mm]
&=(m+1) \alpha\beta+ {\sqrt {\frac{{{d_1}{d_2}({d_1} - 1)({d_2} - 1)}}{4}} }-\alpha\beta\nonumber\\[1mm]
&={\left\| {\begin{array}{*{20}{c}}
   {\alpha \beta {E_{m \times m
   }}} & {\beta \sqrt {\frac{{{d_2}({d_2} - 1)}}{2}} E_{m\times 1}}  \\[1mm]
   {\alpha \sqrt {\frac{{{d_1}({d_1} - 1)}}{2}} E_{1\times m}} & {\sqrt {\frac{{{d_1}{d_2}({d_1} - 1)({d_2} - 1)}}{4}} }  \nonumber\\[1mm]
\end{array}} \right\|_{\text{tr}}}\\
&=\frac{1}{2}\sqrt{(2m\beta^2+d_1^2-d_1)(2m\alpha^2+d_2^2-d_2)},
\end{align}
where the equality (\ref{eq:equality}) has been used in the first and fifth equalities, and, in the third and fourth equalities, we have employed the fact that the trace norm of a Hermitian positive semidefinite matrix is equal to its trace.\hfill \rule{1ex}{1ex}

{\sf Proof of Theorem 2.}~  Without loss of generality, we assume that
\begin{align}
&A=\{q_1,\cdots,q_M\},~q_1<\cdots<q_M,\\
&\bar{A}=\{q_{M+1},\cdots,q_N\},~q_{M+1}<\cdots<q_N.
\end{align}
Since $\rho$ is fully separable, then from \cite{Hassan} there exist vectors $u_i^{(k)}\in \mathbb{R}^{d_k^2-1}$ such that
\begin{align}
\mathcal{W}_{\alpha_1,\cdots,\alpha_N}^{(m)}(\rho)=\sum\limits_i p_i \left( {\begin{array}{*{20}{c}}
   {{\alpha _1}{E_{m \times 1}}}  \\[1mm]
   {u_i^{(1)}}  \\
\end{array}} \right) \otimes  \cdots  \otimes \left( {\begin{array}{*{20}{c}}
   {{\alpha _N}{E_{m \times 1}}}  \\[1mm]
   {u_i^{(N)}}  \\
\end{array}} \right),
\end{align}
where
\begin{equation}
||u_i^{(k)}||_{\text{2}}=\sqrt{\frac{d_k(d_k-1)}{2}}.
\end{equation}
Thus,
\begin{align}
\left\|\left(\mathcal{W}_{\alpha_1,\cdots,\alpha_N}^{(m)}(\rho)\right)^{A|\bar A}\right\|_{\text{tr}}&=\left\|\sum\limits_i p_i  \mathop \otimes \limits_{l=1}^M\left(\begin{array}{*{20}{c}}
   {{\alpha _{q_l}}{E_{m \times 1}}}  \\
   {u_i^{(q_l)}}  \\
\end{array}\right) \mathop \otimes \limits_{p=M+1}^N \left(\begin{array}{*{20}{c}}
   {{\alpha _{q_p}}{E_{m \times 1}}}  \\
   {u_i^{(q_p)}}  \\
\end{array}\right)^t\right\|_{\text{tr}}\nonumber\\[1mm]
&\le\sum\limits_i p_i \left\| \mathop \otimes \limits_{l=1}^M\left(\begin{array}{*{20}{c}}
   {{\alpha _{q_l}}{E_{m \times 1}}}  \\
   {u_i^{(q_l)}}  \\
\end{array}\right) \mathop \otimes \limits_{p=M+1}^N \left(\begin{array}{*{20}{c}}
   {{\alpha _{q_p}}{E_{m \times 1}}}  \\
   {u_i^{(q_p)}}  \\
\end{array}\right)^t\right\|_{\text{tr}}\nonumber\\[1mm]
&=\sum\limits_i p_i \left\| \mathop\otimes\limits_{l=1}^M  \left(\begin{array}{*{20}{c}}
   {{\alpha _{q_l}}{E_{m \times 1}}}  \\
   {u_i^{(q_l)}}  \\
\end{array}\right) \right\|_{\text{2}}\left\|\mathop \otimes\limits_{p=M+1}^N \left(\begin{array}{*{20}{c}}
   {{\alpha _{q_p}}{E_{m \times 1}}}  \\
   {u_i^{(q_p)}}  \\
\end{array}\right)\right\|_{\text{2}}\nonumber\\[1mm]
&=\mathop \Pi \limits_{k = 1}^N
\sqrt{\frac{1}{2}(2m\alpha_k^2+d_k^2-d_k)},
\end{align}
where we have used the equality (\ref{eq:equality}).\hfill \rule{1ex}{1ex}

\newpage
\bigskip
\noindent{\sf Acknowledgements}

\noindent This work is supported by the Fundamental Research Funds for the Central
Universities (No. 15CX05062A, No. 15CX02075A), and the Project-sponsored by SRF for ROCS, SEM.
We are grateful to the referee and the editor
 for their helpful suggestions to improve the quality of this paper.

\bigskip
\noindent{\sf Author contributions}

\noindent  Shen S.Q., Li M. and Fei S.M. wrote the main manuscript text. Yu J. computed the examples. All authors
reviewed the manuscript.

\bigskip
\noindent{\sf Additional Information}

\noindent Competing Financial Interests: The authors declare no competing financial interests.

\end{document}